\documentclass[amsmath,amssymb,twocolumn,aps,pra,floatfix]{revtex4}
\usepackage{graphicx,amsmath,epsfig,epstopdf}
\usepackage{dcolumn}
\usepackage{bm}

\newcommand{\bra}[1]{\ensuremath{\left< #1 \right|}}
\newcommand{\ket}[1]{\ensuremath{\left| #1 \right>}}

\newcommand{\e}{\mbox{e}}

\newcommand{\aaa}{\alpha}
\newcommand{\asd}{\begin{equation}}
\newcommand{\dsa}{\end{equation}}
\usepackage{mathrsfs}
\begin{document}

\title{Violation of Bell's inequalities with pre-amplified homodyne detection}%

\author{G. Torlai$^{1,2}$, G. McKeown$^2$, P. Marek$^3$, R. Filip$^3$, H. Jeong$^{4,5}$, M. Paternostro$^{2}$, and G. De Chiara$^{2}$}
\address{
$^{1}$Department of Physics, Ludwig Maximilians Universit\"at, Schellingstra\ss e 4 80799, Munich, Germany\\
$^{2}$Centre for Theoretical Atomic, Molecular and Optical Physics, School of Mathematics and Physics, Queen's University, Belfast BT7 1NN, United Kingdom\\
$^3$Department of Optics, Palack\'y University, 17. listopadu 1192/12, 77207 Olomouc, Czech Republic\\
$^4$Center for Macroscopic Quantum Control, Department of Physics and Astronomy, Seoul National University, Seoul, 151-742, Korea\\
$^5$Centre for Quantum Computation and Communication Technology, School of Mathematics and Physics, University of Queensland, Brisbane, Queensland 4072, Australia  
}

\date{\today}

\begin{abstract}
We show that the use of probabilistic noiseless amplification in entangled coherent state-based schemes for the test of quantum non locality provides substantial advantages. The threshold amplitude to falsify a Bell-CHSH non locality test, in fact, is significantly reduced when amplification is embedded into the test itself. Such a beneficial effect holds also in the presence of detection-inefficiency. Our study helps in affirming noiseless amplification as a valuable tool for coherent information processing and the generation of strongly non classical states of bosonic systems.
\end{abstract}
\pacs{}
\maketitle

It is well known that entangled two-mode states endowed with a Gaussian Wigner function~\cite{wigner} and subjected to Gaussian phase-space measurements are unable to reveal any non local feature. This point was originally used by Bell to conjecture that the (non normalized) entangled Einstein-Podolski-Rosen state $\int^\infty_{-\infty}dx\ket{x,x+x_0}$~\cite{EPR} (with $\ket{x}$ and $\ket{x+x_0}$ two position eigenstates of a harmonic oscillator), whose Wigner function is positive in the whole phase space, would not falsify any local hidden variable model~\cite{Bell}. 
However, Banaszek and W\'odkiewicz later devised a phase-space approach based on the statistics gathered from the measurement of photon parity operators~\cite{banaszek}, demonstrating the key role played by non-Gaussianity in the revelation of the non local feature of entangled two-mode states. This sort of approach finds its complement in non locality tests performed using Gaussian operations and measurements on non-Gaussian states, such as entangled coherent states (ECSs)~\cite{sanders}, or de-Gaussified two-mode states achieved by resorting to photon subtraction (photon addition)~\cite{olivares,garciapatron,jeong2008,Nha,Park}. On the other hand, recently it has been shown that the combination of Gaussian and non Gaussian measurements can lead to significant violations of local realistic models using continuous-variable systems~\cite{cavalcanti}.

In particular, the non-local nature of an ECS has been extensively studied in past years, addressing tests based on effective pseudo-spin operators, photon parity operators,  effective rotations, and dichotomized homodyne measurements, even in the presence of decoherence~\cite{wilson2002,jeong2003,radim2001,stobinska2007,jeong2009}. The latter approaches have been used for the violation of Bell-CHSH inequality~\cite{clauser1969} by states having a very large thermal occupation number~\cite{jeong2006}, thus showing the possibility to reveal their non classical character even under mechanisms that, naively, would be expected to wash out any quantumness. 
A conspicuous feature of ECS-based tests using homodyne measurements is that the violation of a Bell-CHSH inequality occurs only for coherent-state components having amplitude larger than a given threshold. Under realistic conditions, the threshold is typically determined by the operative conditions (detection inefficiencies and purity of the state resource, among other factors) under which the test is run. In light of the experimental difficulties encountered in the generation of ECS of large-amplitude components~\cite{Ourjoumtsev}, it is clearly desirable to identify viable strategies for the falsification of local realistic theories with lower amplitude thresholds, so as to ease the experimental efforts required for such an important task. 

In this paper we report a test of local realism for ECS of light having an arbitrarily small amplitude, supplemented by the application of local noiseless amplification to the components of the system, after the implementation of the necessary local operations that are part of the Bell test~\cite{zavatta2011,xiang2009}. By increasing the amplitude of the coherent-state components without amplifying the quantum fluctuations, we show that the maximal violation of the Bell inequality can be approached. The threshold for the violation of the CHSH inequality can be considerably lowered, thus realising the mechanism sought above. 

An important point needs to be addressed here. In the experimental scenario where noiseless amplification can only be implemented as a probabilistic heralded process, the implementation of the amplification stage {\it after} the local operations makes the detection of the correct amplified ECS state a probabilistic event which, in turns, opens up a loophole in the test to be run. In fact, this implies that only the cases where the amplification operation is successful for both involved parties of the state at hand should then be considered for the non locality test~\cite{jacobberry}. Such postselection step requires us to invoke a fair-sampling assumption, similarly to what is done for the well-known detection loophole in Bell-CHSH inequalities, an assumption that we will maintain throughout the paper.  As a further result, we mention that the inclusion of noiseless amplification is also beneficial when considering the resilience to key sources of imperfections, such as inefficient measuring apparatuses. The relation between amplification and the detection loophole has been discussed and experimentally shown in~\cite{pomarico}.

We can also treat the Bell inequality violation as an advanced entanglement witness independent of assumed quantum mechanical descriptions of the employed states and measurements \cite{entanglement}. The probabilistic nature of the test is not limiting for this goal---the operations are still local and any detection of entanglement safely implies entanglement in the original state. Thus we demonstrate that, despite its unavoidable probabilistic features, the noiseless amplifier can be useful in state detection, as well as in state preparation. 

The remainder of this paper is organised as follows. In Sec.~\ref{inizio} we gradually introduce the effects of local amplification on the protocol for the violation of Bell-CHSH inequalities with ECS, local rotations, and dichotomic homodyne measurements. We first address the non physical case of ideal noiseless amplification, providing the rationale for our proposal. We  show that the threshold value of  the coherent-state amplitudes for the violation of a Bell-CHSH inequality decreases with the amplification gain. We then turn to an experimentally implementable approximation of the full amplifier, demonstrating that the predicted effect persists even at the lowest significant order (with respect to the gain) in the series expansion of the amplification operator. In Sec.~\ref{secineff} we include the influences of inefficient homodyne detection and the modification to the behavior of the Bell-CHSH function induced by the use of a series of physical operations that, for coherent states of large amplitudes, approximate well the effects of the local rotations. We show that the amplification is effective in reducing the threshold amplitude even under such unfavourable conditions.
Finally, in Sec.~\ref{conc} we draw our conclusions and provide an outlook for future developments along the lines of this paper. 

Our work strengthens the role of noiseless local amplification in coherent quantum information processing, showing its usefulness in the design of tests for the revelation of non classicality in important classes of entangled states.

\section{Bell-CHSH inequality with locally amplified ECS}
\label{inizio}

\subsection{Full amplification}
\label{fullamp}
We consider the unnormalised ECS 
$\ket{\textrm{ECS}_+(\alpha)}=\ket{\aaa,\aaa}+\ket{-\aaa,-\aaa}$
with $\ket{\alpha}$ a coherent state of amplitude $\alpha\in\mathbb{C}$. It is well known that for even moderately large values of $\alpha$, we have $\langle\alpha|-\alpha\rangle\simeq{0}$, which entails the fact that, upon proper normalization, $\ket{\textrm{ECS}_+(\alpha)}$ carries up to a full ebit of entanglement for $\alpha\gtrsim{1}$.  On the other hand, for $\alpha\ll 1$ the state approaches the unnormalized state $|00\rangle+\alpha^2|11\rangle$ in the space spanned by the Fock states $\{|0\rangle$,$|1\rangle\}$, which can also violate a
Bell inequality despite its weak degree of entanglement. However, the entanglement is quite particle type, due to the single-excitation Fock state-based decomposition above. 

\begin{figure}[b]
\centering
\includegraphics[width=85mm]{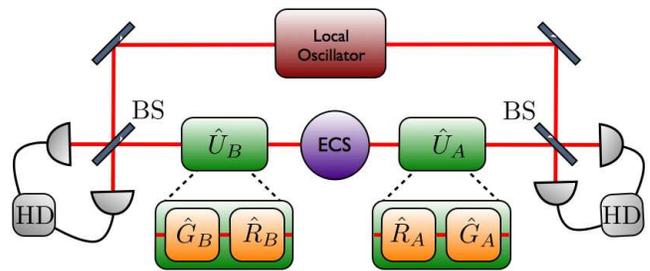}
\caption{(Color online) Scheme for the violation of the CHSH inequality with amplified entangled coherent state. We show the source of ECS states, the local oscillator (LO) needed for homodyne measurements, and  the decomposition of the local unitary transformations $\hat{U}_j$ given in terms of the rotations $\hat R(\theta_j)$ and local amplification $\hat G_j~(j=A,B)$. We also show the symbols for beamsplitters and homodyners.}
\label{schema}
\end{figure}

Following the proposal put forward in Ref.~\cite{stobinska2007}, the non local nature of ECSs can be tested by means of local operations, implemented by cascading linear and non-linear transformations, and dichotomized homodyne measurements. We modify such earlier schemes by introducing, immediately, the key point of our protocol, which consists of supplementing such local transformations with local amplification stages, along the lines of the scheme shown in Fig.~\ref{schema}.
We thus introduce the local transformations 
\begin{equation}
\hat{U}_j=\hat{G}_j\,\hat{R}_j(\theta_j)~~~~(j=A,B),
\end{equation}
with $\hat R_{j}(\theta_j)$ the local rotations in the space spanned by the quasi orthogonal coherent states $\{\ket{\alpha},\ket{-\alpha}\}$ that have been  discussed in~\cite{stobinska2007} and whose form reads
\begin{equation}
\label{rotazioni}
\begin{aligned}
\hat{R}_j(\theta_j){\bm v}_j
=
\begin{pmatrix}
\cos\theta_j&\sin\theta_j\\
\sin\theta_j&-\cos\theta_j
\end{pmatrix}{\bm v}_j.~~(j=A,B).
\end{aligned}
\end{equation}
Here, ${\bm v}_j=(\ket{\alpha}_j\,\ket{-\alpha}_j)^T$ is the vector of coherent-state components for mode $j$.
The other transformation in our scheme is the local noiseless amplification described by the operator $\hat{G}_j=\exp[{(g-1)\hat{a}_j^\dagger\hat{a}_j}]$~\cite{fiurasek2009}, where $g\ge 1$ is the gain of the amplifier and $\hat{a}_j$ and $\hat{a}_j^\dagger$ are the bosonic annihilation and creation operators for mode $j$. For now, we retain the full form of the amplification operator to illustrate, in a clear-cut way, the working principle of our proposal. It should be noticed that in Ref.~\cite{bask} the use of local amplification preceding the local rotations has been discussed for Bell tests purposes, a case that reduces to the offline preparation of a locally amplified resource and does not require any fair-sampling assumption.

\begin{figure}[t]
\includegraphics[width=0.85\columnwidth]{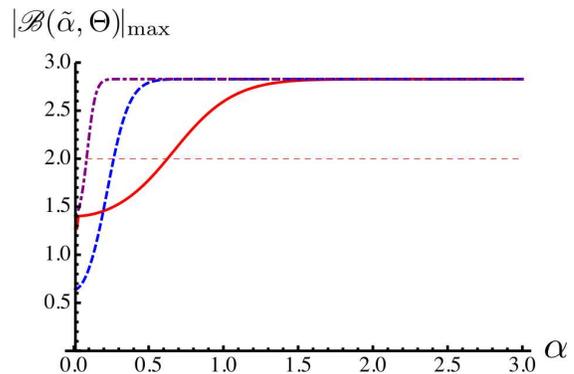}
\caption{(Color online) Bell-CHSH function $\mathscr{B}(\tilde\alpha,\Theta)$, optimized over the set of rotations angles $\Theta$, plotted against $\alpha$ for $g=1$ (solid red line), $g=2$ (blue dashed line) and $g=3$ (purple dot-dashed line). The light straight line marks the local realistic bound.}
\label{tuttoamplificato}
\end{figure}

In what follows, we will retain only the cases where bilateral local amplification is successfully performed. Let us consider the effect of $\hat U_A\otimes\hat U_B$ on the ECS $\ket{\textrm{ECS}_+(\alpha)}$. As $\hat G_j\ket{\alpha}_j=\ket{\tilde{\alpha}}_j$ with $\tilde{\alpha}=\alpha e^{g-1}$, it is straightforward to show that
\begin{equation}
\label{intuitivo}
\begin{aligned}
\ket{\psi_f}&={\cal N}(\hat U_A\otimes\hat U_B)\ket{\textrm{ECS}_+(\alpha)}\\
&={\cal N}\left\{\cos[2(\theta_B-\theta_A)]\ket{\textrm{ECS}_+(\tilde\alpha)}\right.\\
&\left.+\sin[2(\theta_B-\theta_A)]\ket{\textrm{ECS}'_-(\tilde\alpha)}\right\}
\end{aligned}
\end{equation}
with ${\cal N}$ the normalization factor
\begin{equation}
{\cal N}=\left(2+2\nu e^{-4\tilde\alpha^2}\right)^{-1/2},
\end{equation}
$\nu=\cos[2(\theta_A-\theta_B)]$, and where we have introduced the unnormalized ECS $\ket{\textrm{ECS}'_-(\alpha)}=\ket{\alpha,-\alpha}_{AB}-\ket{-\alpha,\alpha}_{AB}$. Eq.~(\ref{intuitivo}) has the very same structure that would be taken by $\ket{\textrm{ECS}_+(\alpha)}$ upon bi-local rotation and no amplification~\cite{stobinska2007}, the only change being the actual amplitude of the coherent-state components. In turn, this implies that, upon application of the proposal for Bell-CHSH test discussed in~\cite{stobinska2007,jeong2009}, which is based on dichotomized homodyne measurements performed on modes $A$ and $B$, we get  the following expression for the correlation function between measurement outcomes following the rotation of the modes' state by $\theta_A$ and $\theta_B$ respectively
\begin{equation}
\mathscr{C}(\tilde\alpha,\theta_A,\theta_B)=\frac{\textrm{Erf}^2[\sqrt2\tilde\alpha]\nu}{1+\nu e^{-4\tilde\alpha^2}}
\end{equation}
with ${\textrm{Erf}}[\cdot]$ the Error function. In this framework, the Bell-CHSH function is written as 
\begin{equation}
\begin{aligned}
\mathscr{B}(\tilde\aaa,\Theta)&=\mathscr{C}(\tilde\aaa,\theta_{A1},\theta_{B1})+\mathscr{C}(\tilde\aaa,\theta_{A1},\theta_{B2})\\
&+\mathscr{C}(\tilde\aaa,\theta_{A2},\theta_{B1})-\mathscr{C}(\tilde\aaa,\theta_{A2},\theta_{B2}),
\end{aligned}
\end{equation}
where $\Theta=\{\theta_{A1},\theta_{A2},\theta_{B1},\theta_{B2}\}$ is a set of rotations angles. Local realistic theories impose the bound $|\mathscr{B}|\le{2}$. Quantum mechanically, this inequality can be violated using ECSs, the set of rotations in Eq.~(\ref{rotazioni}) and dichotomic homodyne detection. From this analysis it is clear that, by calling $\overline{\alpha}$ the amplitude of the coherent-state components at which the Bell-CHSH inequality is first violated and having prepared $\ket{\textrm{ECS}_+(\alpha_a)}$ with $\alpha_a\ll\overline{\alpha}$, we can get $\mathscr{B}>2$ using an appropriate gain, according to the relation
\begin{equation}
g\ge1+\ln(\overline{\alpha}/\alpha_a).
\end{equation}
The behavior of $\mathscr{B}$ against $\alpha$ and for a set of values of the gain is shown in Fig.~\ref{tuttoamplificato}, which demonstrates the quick saturation of the Bell-CHSH function to the Csirel'son bound $2\sqrt2$ and the reduction (exponential with the value of the gain $g$) in the threshold amplitude for the violation of the inequality. 

\subsection{Effective amplification}
It is well known that the unbound nature of $\hat G_j$ makes the transformation $\ket{\alpha}\to\ket{\alpha e^{g-1}}$ unphysical and, as remarked in the previous section, implementable only probabilistically. The realization of noiseless amplification has been at the centre of an intense theoretical and experimental activity~\cite{zavatta2011,fiurasek2009,ralph2009,xiang2009,ferreyrol2010,petrsugg1}. The role of noiseless amplification in quantum information processing and quantum communication has been addressed in an ample variety of ways~\cite{petrsugg1,gisin,bask}. For weak coherent states and small values of gain, the amplification operator can be expanded to the first order in $g$ as~\cite{zavatta2011}
\begin{equation}
\label{approx}
\hat{G}_j\simeq \hat{\openone}+(g-1)\hat{a}_j^\dagger\hat{a}_j=(g-2)\hat{a}^\dag_j\hat{a}_j+\hat{a}_j\hat{a}_j^\dag.
\end{equation}
The amplification thus results in the application of a weighted coherent superposition of the operators $\hat{a}^\dag_j\hat{a}_j$ and $\hat{a}_j\hat{a}_j^\dag$. Both photon-subtraction and addition operations have already been realized experimentally for arbitrary states of light~\cite{zavatta2005/8}. A general superposition of these two operators can be experimentally engineered with a suitable configuration of stimulated parametric down-conversion and linear optics elements and with only a negligible contribution from multiphoton events~\cite{zavatta2009}. 

A remark is due at this stage. 
When Eq.~(\ref{approx}) is used together with the local operations discussed in Sec.~\ref{fullamp} and dichotomized homodyne measurements, the actual ordering of the amplification and rotation stages is key to the success of the overall protocol. In particular, it takes a straightforward, albeit lengthy calculation to show that, when the amplification (with $g\ll{1}$) precedes the bilocal rotations, no advantage with respect to the no-amplification version of the scheme is achieved. Indeed, the state resulting from the application of the operator $\hat{U}'_{A}\otimes\hat{U}'_{B}$ [with $\hat G_j$ approximated as in Eq.~(\ref{approx}) and only the cases of successful bilateral amplification being retained] reads 
\begin{equation}
\label{nouse}
\begin{aligned}
\ket{\psi'_f}&={\cal N}'(\hat{U}'_{A}\otimes\hat{U}'_{B})\ket{\textrm{ECS}_+(\alpha)}\\
&\simeq{\cal N}[\cos[2(\theta_B-\theta_A)]\ket{\textrm{ECS}_+(\alpha)}\\
&+\sin[2(\theta_B-\theta_A)]\ket{\textrm{ECS}'_-(\alpha)}],
\end{aligned}
\end{equation}
which bears no dependence on the amplification gain. Differently, we will prove in what follows that amplification following local rotations indeed results in a more advantageous resource that exhibits features similar to those of the fully amplified state in Eq.~(\ref{intuitivo}). We thus describe the protocol for the construction of the Bell-CHSH function resulting from the application of the $\hat U_j$'s onto $\ket{\textrm{ECS}_+(\alpha)}$ and dichotomized homodyne measurements. This demonstrates that noiseless amplification is important to fullfill the demanding task at the core of this paper. 
 
The initial state $\ket{\textrm{ECS}_+(\alpha)}$ is correspondingly transformed into $\ket{\psi_f}=(\hat{U}_A\otimes\hat{U}_B)\ket{\textrm{ECS}_+(\alpha)}$ and measured via homodyne detection. Taking $\aaa\in\mathbb{R}$ without loss of generality, the joint probability amplitude to get homodyne signals $x_A$ and $x_B$ at sites $A$ and $B$, respectively, is  
\begin{equation}
\label{correide}
C_{g}(x_A,x_B,\theta_A,\theta_B)
=\sum_{\gamma=\pm\aaa}\,\Gamma_\gamma^{g}(x_A,\theta_A)\Gamma_\gamma^{g}(x_B,\theta_B),
\end{equation} 
where 
$\Gamma_{\pm\alpha}^{g}(x_j,\theta_j)={}_j\!\bra{x}\hat{U}_j\ket{\pm\alpha}_j$ and $\ket{x}_j$ is an eigenstate of the quadrature operator $\hat{x}_j=(\hat a^\dag_j+\hat a_j)/2$. 
An explicit calculation gives us
\begin{equation}
\Gamma_{\pm\aaa}^{g}(x_j,\theta_j)=\frac{1}{\sqrt[4]\pi}[\xi_{\mp\aaa}(x_j)\sin\theta_j\pm\xi_{\pm\aaa}(x_j)\cos\theta_j],
\end{equation}
where we have introduced the functions $\xi_{\pm\aaa}(y)=\e^{-(y\mp\aaa)^2}[1+(g-1)(\pm2\aaa y-\aaa^2)]~(y=x_A,x_B)$. To construct the Bell function the continuous variables must be dichotomized. This is done by assigning a value +1 to a homodyne measurement larger than 0 and -1 otherwise, constructing in this way a set of dichotomic observables. The joint probabilities of the measurement outcomes are
\begin{equation}
P_{kl}^g(\theta_A,\theta_B)=\frac{1}{K}\int_{\Omega_k}dx_A\int_{\Omega_l}dx_B\,|C_g(x_A,x_B,\theta_A,\theta_B)|^2,
\end{equation}
where $k$,$l=\pm$ correspond to the bilateral measurement outcomes $\pm1$, $\Omega_+=[0,\infty]$, $\Omega_-=[-\infty,0]$ and 
$K$ is a normalization constant. 
The Bell-CHSH function is then
\begin{equation}
\begin{aligned}
B^g(\aaa,\Theta)&=\mathscr{C}^g(\aaa,\theta_{A1},\theta_{B1})+\mathscr{C}^g(\aaa,\theta_{A1},\theta_{B2})\\
&+\mathscr{C}^g(\aaa,\theta_{A2},\theta_{B1})-\mathscr{C}^g(\aaa,\theta_{A2},\theta_{B2}),
\end{aligned}
\end{equation}
 with the correlation function
\begin{equation}
\begin{aligned}
&\mathscr{C}^g(\aaa,\theta_A,\theta_B)=\sum_{k=\pm}P_{kk}^g(\theta_A,\theta_B)-\sum_{k\ne l=\pm}P_{kl}^g(\theta_A,\theta_B)\\
&=\frac{\sqrt{\mu_\aaa}\,\nu\,\mbox{Erf}[\sqrt{2}\aaa]}{\sqrt{\pi}(\mu_\aaa+\nu)^2}\left\{4\sqrt{2}\alpha(g-1)(\mu_\aaa+\nu){+}\sqrt{\pi\mu_\aaa}\mbox{Erf}[\sqrt{2}\aaa]\right.\\
&\times[\mu_\aaa+(1+8(g-1)\aaa^2)\nu]\Big\}
\end{aligned}
\end{equation}
and $\mu_\aaa=\exp[{4\aaa^2}]$. 
\begin{figure}[b!]
\includegraphics[width=0.9\columnwidth]{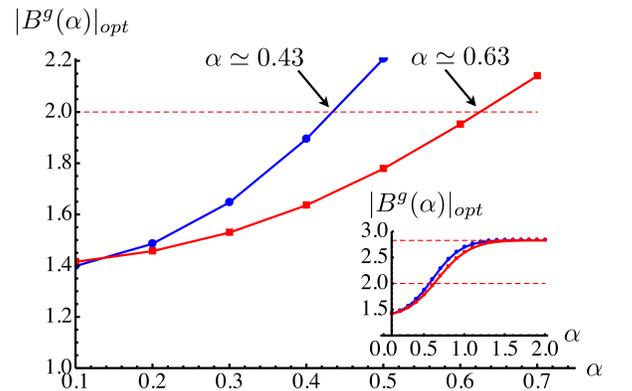}
\caption{(Color online) Bell function (optimized numerically over the set of rotation angles $\Theta$) plotted against the coherent-state amplitude $\alpha$ for $g=1.0$ (red curve) and $g=1.4$ (blue curve). Inset: Same as in the main panel but for $g=1.0$ (red curve) and $g=1.1$ (blue curve) and $\alpha\in[0,2]$. Even small increases in the gain factor result in noticeable reductions of the threshold for the violation of the Bell-CHSH inequality.}
\label{fig2}
\end{figure}
While a local realistic description of the entangled coherent state in the presence of the ideal local rotations and without amplification is not possible for $\aaa\gtrsim0.63$, for a state locally amplified by $g=1.1$ such threshold is lowered to 0.57 [cf. inset of Fig.~\ref{fig2}]. 

Further reductions of the threshold value of $\alpha$ can be obtained by increasing the gain, still remaining within the limits of validity of the second-order expansion within which our calculations have been performed. For instance, 
in the main panel of Fig.~\ref{fig2} we show the Bell function, optimised numerically over $\Theta$, for $g=1$ (red curve) and $g=1.4$ (blue curve), plotted against the coherent-state amplitude $\alpha$. The value of $\aaa$ at which the Bell-CHSH inequality is first violated when the state is locally amplified goes down to 0.43, an approximately $30\%$ reduction in the value corresponding to no local amplification. In this case the inaccuracy due to the second order expansion in $g$ is about $2\times10^{-3}$. As an example, we report the value of the optimized Bell's function without amplification for $\aaa=0.7$ which is $B_{id}^1(0.7,\Theta_{0.7})\simeq2.14$, and compare it to $B_{id}^{1.4}(0.7,\Theta^g_{0.7})\simeq2.76$, which corresponds to $g=1.4$. 
We can see that, already at $\aaa=0.7$, the Bell's function is almost saturated. 

\section{Inefficient homodyne detectors and effective rotations}
\label{secineff}
In this section we show the effect that inefficient detectors have on the behavior of the Bell function. Moreover, we replace the idealized local rotations in Eq.~(\ref{rotazioni}) with a cascade of local unitary operations whose resulting effect on a single mode is to approximate $\hat R_j(\theta_j)$. 
As shown in Refs.~\cite{stobinska2007,jeong2009}, both the detection inefficiency and the replacement of the idealized rotations with effective ones increase the threshold value of $\aaa$ for the violation of the Bell-CHSH inequality. 

Let us start with the analysis of non ideal homodyne detectors, each being modelled as a perfect detector preceded by beam splitters of transmittivity $\eta$. The latter mixes mode $j$ to an ancillary mode $\hat{a}_{j}~(j=A,B)$ prepared in the vacuum state. At the output port of the beam splitter, the reduced state of mode $j$ (after tracing out the corresponding ancilla) describes the signal detected by a homodyner of efficiency $\eta$.

\begin{figure}[t]
\centering
\includegraphics[width=0.9\columnwidth]{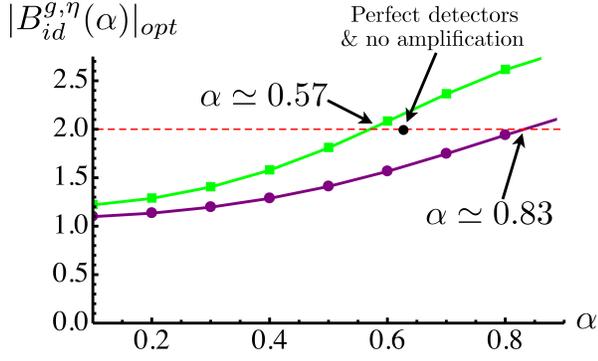}
\caption{(Color online) Numerically optimized Bell's function plotted against the amplitude of the coherent states with detection inefficiencies. The black point indicates the value of $\aaa$ for which the violation occurs with perfect detectors ($\eta=1$) and no amplification. Setting $\eta=0.9$ we obtained the purple curve for $g=1.0$ and the green curve for  $g=1.4$.}
\label{ineff}
\end{figure}

By proceeding along the lines of the calculations described in Sec.~\ref{inizio}, we get the correlation function
\begin{equation}
\begin{aligned}
&\mathscr{C}_{\textrm{d}}^g(\aaa,\theta_A,\theta_B)=\frac{\mu_\aaa\nu\,\e^{-2\kappa_\eta^2}\mbox{Erf}(\sqrt{2}\kappa_\eta)}{[\sqrt{\pi}(\mu_\aaa+\nu)^2]}\left\{4\sqrt{2}\kappa_\eta(\mu_\aaa+\nu)\right.\\
&\left.+\sqrt{\pi}\e^{2\kappa_\eta^2}\Big[\mu_\aaa+[1+2(g-1)(4\aaa^2+\kappa_\eta^2)]\nu\,\mbox{Erf}(\sqrt{2}\kappa_\eta)\Big] \right\}\notag
\end{aligned}
\end{equation}
with $\kappa_\eta=\eta\aaa$. 
In Fig.~\ref{ineff} we compare the optimized Bell function for no gain and detection efficiency $\eta=0.9$ to what is obtained by introducing the local amplification stages (with $g=1.4$) and for the same value of $\eta$. The amplified ECS violates the Bell-CHSH inequality for smaller values of $\alpha$ than the non amplified state affected by the same degree of detection inefficiency. It also overcomes the performance of the Bell function for no amplification and ideal homodyne detectors.  

\begin{figure}[!t]
\includegraphics[width=0.9\columnwidth]{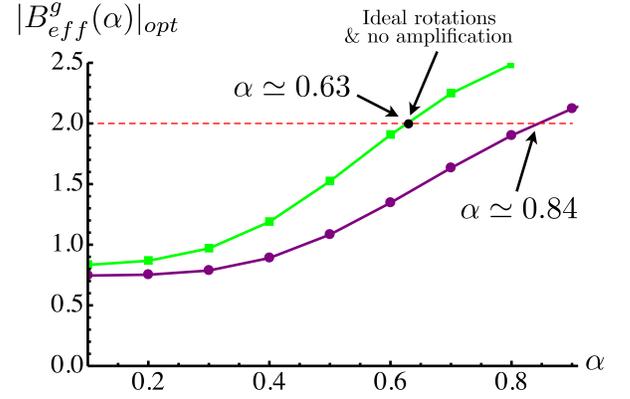}
\caption{(Color online) Numerically optimized Bell's function plotted against the amplitude of the coherent states with effective rotations for $g=1.0$ (purple curve) and $g=1.3$ (green curve). The black point represents the value of $\aaa$ for which the violation occurs with ideal rotations and no amplification.}
\label{effrot}
\end{figure}

We now pass to the construction of the correlation function resulting from the use of the operations approximating the local rotation operators on each mode of our system. In order to simplify our mathematical approach, from here on we will only consider ideal detectors, the extension to imperfect ones being performed following the lines sketched above. Equation~(\ref{rotazioni}) is well approximated by the cascade of the transformation resulting from the self-Kerr Hamiltonian $\hat{H}_j=\hbar\Omega(\hat{a}_j^\dagger\hat{a}_j)^2$ and a phase-space displacement by an appropriate amplitude according to the overall expression
\begin{equation}
\hat{V}_j(\theta_j)=\e^{i\pi(\hat{a}_j^\dag\hat{a}_j)^2}\hat{D}_j({i\theta_j}/{\aaa})\,\e^{i\pi(\hat{a}_j^\dag\hat{a}_j)^2}.
\end{equation}
When applied to the components of the set of quasi-orthogonal states $\{\ket{\alpha}_j,\ket{-\alpha}_j\}$, this leads to the following set of transformations~\cite{stobinska2007}:
\begin{equation}
\begin{aligned}
\hat{V}_j(\theta_j)\ket{\aaa}_j&=\frac{1}{2}\left[\e^{i\theta_j}(|\aaa+{i\theta_j}/{\aaa}\rangle_j+i|-\aaa-{i\theta_j}/{\aaa}\rangle_j)\right.\\
&\left.+i\e^{i\theta_j}(|-\aaa+{i\theta_j}/{\aaa}\rangle_j+i|\aaa-{i\theta_j}/{\aaa}\rangle_j)\right]\\
\hat{V}_j(\theta_j)\ket{-\aaa}_j&=\frac{1}{2}\left[i\e^{i\theta_j}(|\aaa+{i\theta_j}/{\aaa}\rangle_j+i|-\aaa-{i\theta_j}/{\aaa}\rangle_j)\right.\\
&\left.+\e^{i\theta_j}(|-\aaa+{i\theta_j}/{\aaa}\rangle_j+i|\aaa-{i\theta_j}/{\aaa}\rangle_j)\right].
\end{aligned}
\end{equation}
In order to evaluate the correlation functions upon local rotations and homodyne detection, we replace Eq.~(\ref{correide}) with 
\begin{equation}
C^{\text{eff}}_g(x_A,x_B,\theta_A,\theta_B)=\sum_{\gamma=\pm\aaa}\,\Pi_\gamma^g(x_A,\theta_A)\Pi_\gamma^g(x_B,\theta_B)
\end{equation}
with $\Pi_{\pm\alpha}^g(x_j,\theta_j)=_j\!\!\bra{x_j}\hat{G}_j\hat{V}_j(\theta_j)\ket{\pm\alpha}_j$. We get 
\begin{equation}
\begin{aligned}
\Pi_{\pm\aaa}^g(x_j,\theta_j)&=\mp\frac{i^{{\delta_{\aaa}^{\pm\aaa}}}}{\sqrt[4]{\pi}}\left\{i\e^{i\theta_j}[\xi^+_{\chi_+}(x_j,\theta_j)+i\xi^-_{\chi_+}(x_j,\theta_j)]\right.\\
&\left.\mp\e^{-i\theta_j}[\xi^-_{\chi_-}(x_j,\theta_j)+i\xi^+_{\chi_-}(x_j,\theta_j)]\right\},
\end{aligned}
\end{equation}
where we have introduced $\chi^j_\pm=\aaa\pm\frac{i\theta_j}{\aaa}$ and 
\begin{equation}
\xi^\pm_{\chi_\pm}(x_j,\theta_j)=\e^{-(x_j\mp\chi^j_\pm)^2}[1+(g-1)(\pm2\chi^j_\pm x_j-{\chi^{j2}_\pm})].
\end{equation}
Fig.~\ref{effrot} 
shows the optimized Bell's function with effective rotations for $g=1.0$ (purple curve) and $g=1.3$ (green curve). In this case, the threshold for the violation of the Bell-CHSH inequality is lowered from $\aaa=0.84$, which is the value achieved using the effective rotations, to $\aaa=0.63$, corresponding to the use of the ideal rotation, perfect homodyne measurements, and no amplification. 

\section{Conclusions and outlook}
\label{conc}

We have discussed the effectiveness of local noiseless amplification in lowering the threshold for the violation of a Bell-CHSH inequality by an ECS. The strategy that we have applied consists of local rotations performed over the two modes of the system followed by local  amplification and dichotomic homodyne measurements, which are known to be effective in revealing the non local properties of ECSs. With the underlying fair-sampling assumption needed by the inherent probabilistic nature of experimental noiseless amplification operations, the advantages of using local amplifiers is evident in a significant reduction of the amplitude of the coherent-state components of the ECS needed to go beyond the bound imposed by local realistic theories. 
It will be very interesting to extend the domain of usefulness of local noiseless amplification for quantum information processing by addressing the violation of a Bell-CHSH inequality through local photon parity measurements performed over entangled Gaussian states, such as a two-mode squeezed vacuum state. Our task is to affirm approximate noiseless amplification as a valid and viable alternative to the use of conditional photo-subtraction for the enhancement of the non locality properties of interesting classes of continuous-variable states.

\acknowledgments
G.T. and G.McK. thank the Centre for Theoretical Atomic, Molecular, and Optical Physics, Queen's University Belfast, and the Department of Optics, Palack\'y University, respectively, for hospitality during various stages of this work. We acknowledge financial support from the UK EPSRC through a Career Acceleration Fellowship and the ``New Directions for EPSRC Research Leaders" initiative (EP/G004759/1) as well as the National Research Foundation of Korea (NRF) grant funded by the Korean Government (No. 2010-0018295). RF and PM acknowledge project P205/12/0577, while P.M. also acknowledges P205/10/P319, of GA \v{C}R.

\end{document}